\documentclass[twocolumn,prl,amsmath,amssymb]{revtex4-1}

\usepackage[pdftex]{graphicx}
\usepackage{dcolumn}
\usepackage{bm}
\usepackage[T1]{fontenc}
\usepackage[utf8]{inputenc}
\usepackage{natbib}
\usepackage{hyperref}
\usepackage{color}
\usepackage{url}

\definecolor{OrangePastel}{RGB}{255,200,31}
\definecolor{GreenPastel}{RGB}{33,219,77}
\definecolor{VioletPastel}{RGB}{200,175,242}
\definecolor{RedPastel}{RGB}{255,125,82}
\definecolor{GreenDarkPastel}{RGB}{33,150,77}
\definecolor{GreenLightPastel}{RGB}{33,255,77}
\definecolor{GreenBabethPastel}{RGB}{209,255,108}
\definecolor{BlueBabethPastel}{RGB}{119,207,255}

\newcommand{\Rey}{R\kern-.05em e}
\newcommand{\Capp}{C\kern-.11em a}
\newcommand{\Bond}{B\kern-.09em o}

{\rule{20pt}{1ex}\endlist}
\let\oldmarginpar\marginpar
\renewcommand\marginpar[1]{\-\oldmarginpar[\raggedleft\footnotesize #1]%
{\raggedright\footnotesize #1}}

\begin{document}
\newcommand{\comment}[1]{
\textcolor{red}{\textbf{\textsf{{\large[}#1{\large]}}}}
}

\title{Role of uncrosslinked chains in droplets dynamics on silicone elastomers}
\author{Aur\'elie Hourlier-Fargette$^{1,2}$}
\author{Arnaud Antkowiak$^{1,3}$}
\author{Antoine Chateauminois$^{4}$}
\author{S\'ebastien Neukirch$^{1}$}
\affiliation{
$^1$ Sorbonne Universit\'es, UPMC Univ Paris 06, CNRS, UMR 7190, Institut Jean Le Rond d'Alembert, F-75005 Paris, France. \\
$^2$ D\'epartement de Physique, \'Ecole Normale Sup\'erieure, CNRS, PSL Research University, F-75005 Paris, France.\\
$^3$ Surface du Verre et Interfaces, UMR 125 CNRS/Saint-Gobain, F-93303 Aubervilliers, France. \\
$^4$ ESPCI \& CNRS, UMR 7615, Laboratoire de Sciences et Ing\'enierie de la Mati\`ere Molle, F-75005 Paris, France. \\}

\date{\today}

\begin{abstract}
We report an unexpected behavior in wetting dynamics on soft silicone substrates: the dynamics of aqueous droplets deposited on vertical plates of such elastomers exhibits two successive speed regimes. This macroscopic observation is found to be closely related to microscopic phenomena occurring at the scale of the polymer network: we show that uncrosslinked chains found in most widely used commercial silicone elastomers are responsible for this surprising behavior.  A direct visualization of the uncrosslinked oligomers collected by water droplets is performed, evidencing that a capillarity-induced phase separation occurs: uncrosslinked oligomers are extracted from the silicone elastomer network by the water-glycerol mixture droplet. The sharp speed change is shown to coincide with an abrupt transition in surface tension of the droplets, when a critical surface concentration in uncrosslinked oligomer chains is reached. We infer that a droplet shifts to a second regime with a faster speed when it is completely covered with a homogeneous oil film.
\end{abstract}

\maketitle

%
%
%
%
%
%
%
%
%
%
\section{Introduction}
%
%
%
%
%
%
%
%
%
Commonly used in industry, silicone elastomers also serve as easy-to-make substrates in various academic research domains. Very popular in elastocapillarity, for both experiments on slender bendable structures \cite{Py2007,Fargette2014} and on thick softer substrates \cite{Carre1996,Style2013,Park2014,Karpitschka2016}, they are even more widely spread in microfluidics, {\em e.g.} for biological cultures in microfluidic channels \cite{Berthier2012, Regehr2009}.
However, drawbacks in the use of these elastomers have been reported, for example absorption issues into the polymer bulk \cite{Mukhopadhyay2007}, or leaching of unreacted oligomers from the polymer network into the microchannel medium \cite{Regehr2009}.
Beyond the interest of understanding droplet dynamics on stiff inclined surfaces \cite{Huh1971, Voinov1976, Cox1986, Dussan1985, Shanahan2000, Le-Grand2005}, the development of soft materials has lead to a growing interest for soft-wetting dynamics. A droplet deposited on a soft substrate is able to induce a wetting ridge at the contact line \cite{Limat2012, Park2014}, resulting in an additional source of dissipation when this triple line is moving \cite{Shanahan2002}.
The recent design of Slippery Lubricant Infused Porous Surfaces also leads to novel wetting dynamics studies \cite{Smith2013}, to understand how droplets slide or roll on a micro- or nano-textured substrate infused with oil.
Soft commercial elastomers like poly(dimethylsiloxane) (PDMS) Sylgard 184 from Dow Corning are known to contain a small fraction of uncrosslinked low-molecular-weight oligomers \cite{Lee2003}, the effects of which on wetting dynamics are not completely understood.
However, it has been shown recently that adhesion of a silica microbead on a soft elastomer gel leads to a phase separation which transforms the classical three-phase contact line into a four-phase contact zone, in which air, silica, liquid silicone, and silicone gel meet \cite{Jensen2015}. 

\begin{figure}[bht]
\noindent\includegraphics[width=8.6cm]{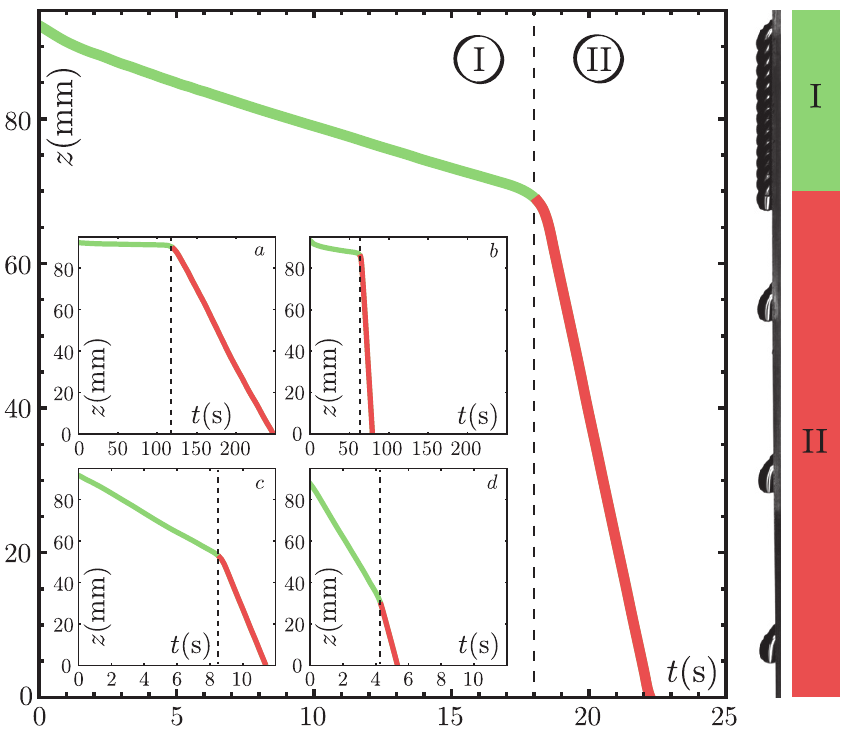}
\caption{A 40\% water - 60\% glycerol mixture droplet of volume 21.5~$\mu$L is deposited with no initial speed on a vertical PDMS surface. Two regimes with two different constant speeds are identified (I and II). 
{\em Right}: Snapshots are taken every 1.28~s and superimposed together. 
{\em Inset}: Same experiment with droplets of volumes 14~$\mu$L (a), 18~$\mu$L (b), 22.5~$\mu$L (c) and 24~$\mu$L (d).}
\label{fig:2regimes}
\end{figure}

Here we show that uncrosslinked chains also play a major role in wetting dynamics in spite of their small amount within the silicone elastomers under consideration: A droplet of water-glycerol mixture deposited on a vertical silicone elastomer surface with no initial speed is shown to exhibit two different regimes characterized by two different constant speeds, as illustrated in Fig.~\ref{fig:2regimes} and in Video 1, Supplementary Information.
We demonstrate the crucial role of uncrosslinked oligomers in this two-regime dynamics, and highlight that the sharp transition between the two speed regimes coincides with a sharp transition in surface tension as the droplet is progressively covered by uncrosslinked chains. Experimental results in the literature \cite{Lee1991} show a correlation between the surface tension and the thickness of a PDMS oil layer on a water bath. We infer from our surface-tension measurements that the two regimes are linked to two different states of the droplet: in the first regime, the droplet is only partially covered with some patches of uncrosslinked oligomers, while in the second regime, the droplet is completely covered by a uniform oil layer, with probably oil wetting ridges at the triple line. We show that the first regime dynamics is set by the competition between the weight of the droplet, capillary pinning forces, and viscous dissipation inside the drop, while the second regime presents some similarities with the dynamics observed in the case of water droplets on Slippery Lubricant Infused Porous Surfaces \cite{Smith2013}.

%
%
%
%
%
%
\section{Experimental methods}
%
%
%
%
%
%
\subsection{Silicone elastomer samples}

Unless otherwise specified, elastomer samples are made of PDMS (Dow Corning Sylgard 184 Elastomer base blended with its curing agent in proportion 10:1 by weight, cured during 2 hours at 60\ensuremath{^\circ}C) molded in 12 cm square Petri dishes to get flat samples with a thickness of a few millimeters. The Young's modulus of our PDMS samples, measured with a Shimadzu testing machine, was found to be $E=1.8\pm0.1$~MPa.
In order to test the generality of the observed phenomenon, some experiments were also carried out using other commercially available silicone elastomers, such as polyvinylsiloxane (PVS, Elite Double 22 Zhermack), RTV EC13 and EC33 polyaddition silicone polymer (Esprit Composite), and PDMS Sylgard 184 cured during 24h at 120\ensuremath{^\circ}C.

\subsection{Droplet deposition and video acquisition techniques}

Droplets of water-glycerol mixtures (deionized water and glycerol 99+\% pure, Acros Organics), which properties are given in Table \ref{tbl:liquids}, are deposited on polymer samples with an electronic micropipette (Sartorius eLINE 5-120 $\mu$L). Surface tensions have been measured with a Kr\"uss K6  manual tensiometer (hanging ring). Videos have been captured with a Hamamatsu Orca Flash digital CMOS camera, with frame rates going from 10 to 1000 frames per second. 

\begin{table}
\begin{ruledtabular}
\begin{tabular}{ccccccc}
Gl. & Wa. & $T$($^{\circ}$C)  & $\mu$ (mPa.s) & $\rho$ (kg.m$^{-3})$ & $\gamma$ (mN.m$^{-1})$\\
\hline
50\% & 50\% & 19.3 $\pm$ 1.0 & 6.2 $\pm$ 0.2 & 1124 $\pm$ 1 & 69.0 $\pm$ 0.5 \\
60\% & 40\% & 25.3 $\pm$ 1.0 & 8.7 $\pm$ 0.2 & 1150 $\pm$ 1 & 68.4 $\pm$ 0.5 \\
65\% & 35\% & 27.0 $\pm$ 1.0 & 11.1 $\pm$ 0.2 & 1164 $\pm$ 1 & 67.9 $\pm$ 0.5 \\
75\% & 25\% & 20.7 $\pm$ 1.0 & 34.1 $\pm$ 0.5 & 1192 $\pm$ 1 & 67.1 $\pm$ 0.5 \\
85\% & 15\% & 27.5 $\pm$ 1.0 & 68.1 $\pm$ 1.0 & 1219 $\pm$ 1 & 66.5 $\pm$ 0.5 \\
\end{tabular}
\end{ruledtabular}
\caption{Mixing ratios (Gl. : glycerol, Wa. : deionized water), temperature $T$ during the experiments, viscosity $\mu$, density $\rho$ and surface tension $\gamma$, measured experimentally for the different liquids used.}
\label{tbl:liquids}
\end{table}

%
%

%
%
%
%
%
%
%
%
\section{Role of uncrosslinked polymer chains}
%
%
%
%
%
%
%
%
%
\subsection{Unexpected dynamics for an aqueous droplet on a vertical elastomer plate}

Fig.~\ref{fig:2regimes} shows the dynamics of a water-glycerol mixture droplet deposited on a vertical PDMS plate. Two distinct regimes characterized by two different constant speeds are evidenced. This two-regime behavior is observed for various droplet volumes (Insets of Fig.~\ref{fig:2regimes}) and various water-glycerol mixing ratios, from pure water to 90\% glycerol mixture.
The same qualitative behavior is also obtained on several commercial silicone elastomers, as shown in Fig.~S1, Supplementary Information. These observations contrast with experiments on treated glass \citep{Le-Grand2005} during which a single constant speed is reached after a short transient. 

Droplet dynamics on a vertical surface usually results from a competition between the weight of the droplet, capillary forces, and viscous dissipation inside the drop.
For a droplet of volume $V$ and density $\rho$ on a vertical surface, the gravitational force is given by:
\begin{equation}
F_g=\rho \, V g \, .
\end{equation}
The descent of the drop begins only if $F_g>F_{cl}$, $F_{cl}$ being the contact-line pinning force:
\begin{equation}
F_{cl}\propto \gamma \, w \, (\cos{\theta_{sr}}-\cos{\theta_{sa}}),
\end{equation}
where $\gamma$ is the surface tension of the liquid-air interface, $w$ the width of the droplet, $\theta_{sr}$ the static receding angle and $\theta_{sa}$ the static advancing angle.
A more precise determination of the onset of motion is given in the literature\cite{Dussan1985}.
Once steady motion is reached, the value of the droplet speed is obtained by balancing the weight $F_g$ with the sum of the contact line pinning force $F_{cl}$ and the viscous dissipation force in the droplet $F_{\mu}$. 
$F_{\mu}$ can be derived using different approaches, depending on the dissipation being located in the bulk or in the receding corner of the droplet \cite{degennes+al:2003,Puthenveettil2013}. The simplest approach yields:
\begin{equation}
F_{\mu}\propto \frac{\mu U \mathcal{S}}{h},
\end{equation}
where $U$ is the droplet speed, $\mu$ the viscosity of the liquid, $\mathcal{S}$ the contact area, and $h$ the droplet height.
For low $\Rey$ viscous drops (with $\Rey= U V^{1/3} \rho / \mu$), assuming that $w=h=\sqrt{\mathcal{S}}=V^{1/3}$ leads to a linear relationship between $\Capp$ and $\Bond$ where $\Capp=\mu U / \gamma$ is the capillary number, and $\Bond=\rho V^{2/3} g / \gamma$ is the Bond number (this relationship has been found to capture the dynamics of droplets even when $\Rey$ is of order one \cite{Le-Grand2005}). This description, as well as more detailed approaches taking into account the details of the viscous dissipation processes in the droplet \cite{Cox1986, Voinov1976, Puthenveettil2013} or in the elastomer \cite{Carre1996,Shanahan2002,Karpitschka2015}, cannot explain the two regimes in our experiments.

Different candidates can be responsible for the sudden change in the droplet speed: droplet shape bistability, modification of the droplet composition due to an interaction with the substrate, sliding to rolling transition, change in shape of the tail of the droplet and pearling transition \cite{Podgorski2001,Ben-Amar2003,Le-Grand2005}. Volumes have been kept under the threshold for which pearling occurs.
Tracking particles in the water-glycerol mixture reveals that, in both regimes, the drop is rolling rather than sliding over the surface, with no significant difference in the motion of the fluid between the two regimes (Video 2, Supplementary Information).\\
Turning the setup upside down after a droplet has slid down allows us to reuse the same droplet and same sample for a second descent and to observe that the behavior of a droplet during the first descent is completely different from its behavior during the following descents. The first descent shows two distinct regimes, whereas the following descents only exhibit one regime, corresponding to the second regime of the first descent (Fig. S2, Supplementary Information).
This observation advocates for a modification of the droplet composition during its first descent.

\subsection{Toluene-treated plates}

When PDMS oligomers are crosslinked to form the polymer network, a few oligomer chains are not incorporated into the network \cite{Lee2003}. A PDMS sample thus contains uncrosslinked low-molecular-weight oligomers, which can be extracted from the bulk PDMS by swelling in a good solvent.
Accordingly, treated PDMS samples are obtained using the following procedure: The samples are fully immersed for one week in a toluene bath which is renewed every day in order to keep the concentration in extracted PDMS oligomers low. Following this extraction, the PDMS substrate is subsequently de-swollen by progressively replacing toluene by a poor solvent of PDMS, namely ethanol. The samples are ultimately dried out in a vacuum oven. An average mass loss of 5\% is measured, which corresponds to the weight fraction of the extracted uncrosslinked chains.

\begin{figure}[bht]
\noindent\includegraphics[width=8.7cm]{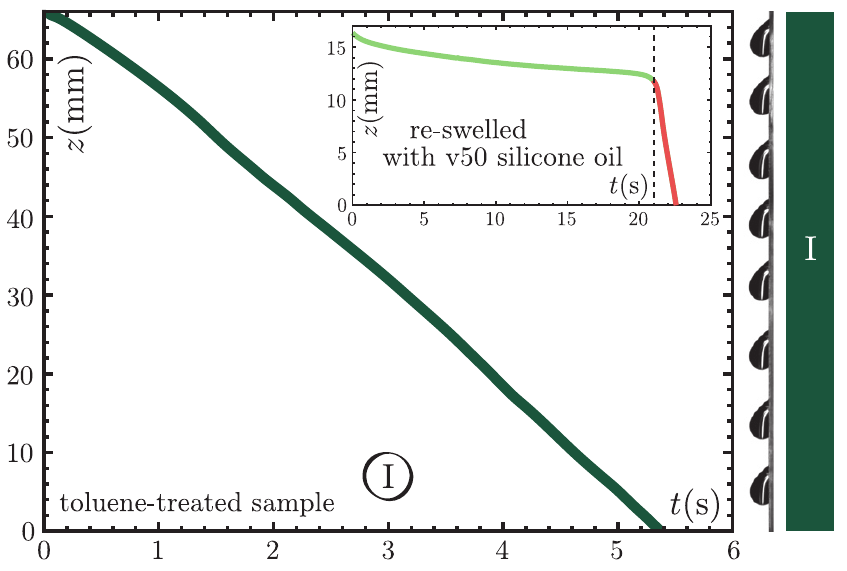}
\caption{A 40\% water - 60\% glycerol mixture droplet of volume 21.5~$\mu$L is deposited with no initial speed on a vertical toluene-treated PDMS surface and a single regime is identified.
{\em Right}: Snapshots are taken every 0.64~s and superimposed together.
{\em Inset}: Droplet dynamics on a toluene-treated PDMS sample re-swelled with commercial v50 PDMS oil. A 50\% water - 50\% glycerol mixture droplet of volume 18~$\mu$L is deposited with no initial speed on a vertical surface and two different regimes with two constant speeds are identified.} 
\label{fig:1regime}
\end{figure}

As illustrated in Fig.~\ref{fig:1regime}, a water-glycerol mixture droplet rolling down on such a treated elastomer only exhibits one speed regime. A one-regime behavior is also observed for various droplet volumes and various water-glycerol mixing ratios, from pure water to 90\% glycerol mixture, deposited on vertical toluene-treated PDMS plates. These results show that the two-regime behavior observed on untreated samples is due to the presence of uncrosslinked oligomer chains.
Reswelling a toluene-washed sample with a commercial PDMS v50 oil (Sigma Aldrich) enables us to recover the two-regime behavior, as shown in the inset in Fig.~\ref{fig:1regime}. This experiment provides an additional evidence of the crucial role played by the free oligomers of the substrate.
Enhanced wetting hysteresis have already been observed in the case of PDMS grafted chains \cite{She1998} but could not explain a two regime behavior. Here, a possible migration of the uncrosslinked chains to the water-air interface is investigated.

In the case of lubricant-infused surfaces, where a water drop rests on top of a thin layer of liquid oil, different wetting configurations are possible depending on the relative intensity of the surface tensions \cite{Smith2013, Schellenberger2015}.
Denoting $\gamma_{wa}$, $\gamma_{wo}$, and $\gamma_{oa}$ the surface tensions of the water-air, water-oil, and oil-air interfaces respectively and introducing the wetting parameter $S=\gamma_{wa}-(\gamma_{wo}+\gamma_{oa})$, there are typically two possible situations for the wetting configuration: If $S>0$ the oil completely covers the droplet and forms a thin film around it, whereas if $S<0$ only a ridge of oil forms at the triple line.  Interesting results about the composition of the uncrosslinked chains likely to migrate at the water-air interface are found in the literature in the context of biological cell cultures in PDMS microchannels: A mass spectroscopy analysis of water aspirated from Sylgard 184 PDMS microchannels has given evidence of the presence of free chains of various lengths, in a continuous range of fewer than 20 to more than 90 dimethylsiloxane units\cite{Regehr2009}. As in the case of a v50 commercial silicone oil, the oil spreading parameter on water is positive, we conjecture the formation of a thin film of liquid PDMS around water-glycerol mixture droplets.

Now in the case where the substrate is a soft PDMS gel, uncrosslinked chains may phase-separate from the core of the gel and act as the thin layer of liquid oil mentioned above -- such a phase separation was reported in the case of adhesion \cite{Jensen2015}.
Here we show that a water droplet can extract uncrosslinked oligomers from the PDMS elastomer, evidencing the existence of capillarity-induced phase separation at the triple line. 
%
This hypothetical oil film on a single droplet is anyhow invisible to the naked eye, but collecting 1500 such droplets in a beaker after their two-regime descent on an untreated PDMS sample results in a direct visualization of oil at the surface of the beaker, as shown in Fig.~\ref{fig:surfacetension}c. The colored zones correspond to thicknesses of the order of light wavelengths, transparent lenses correspond to thicker oil zones, while the grey background corresponds to a thinner zone. The oil film is not homogeneous, in agreement with results in the literature about PDMS oil spreading at the surface of a water bath  \cite{Lee1991}. It is thus difficult to evaluate the quantity of oil extracted by each droplet. A rough estimate, obtained by multiplying the surface of the beaker by a thickness of 500  nm (typical optical wavelength) gives the following result: the order of magnitude of the oil volume collected by each droplet is about $10^5$ times smaller than the typical droplet volume.
The control experiment on a toluene-treated PDMS sample leads to no visible oil at the surface of the beaker.

\subsection{A silicone oligomer coating around a droplet modifies its surface tension}

Surface-tension measurements of PDMS oil chains spread on a flat water-air interface performed in the literature \cite{Lee1991} give some insight into our current work: The surface tension of a water bath covered with PDMS oil chains is shown to be equal to pure water surface tension under a critical surface concentration in PDMS chains, and suddenly decreases above this threshold, to reach a plateau corresponding to the surface pressure of a homogeneous interface. Similar surface tension observations performed with natural oils such as olive or castor oil were already reported more than one century ago by A. Pockels \cite{Pockels1892}, Lord Rayleigh \cite{Rayleigh1899}, and I. Langmuir \cite{Langmuir1917}. 

\begin{figure}[bht]
\noindent\includegraphics[width=8.7cm]{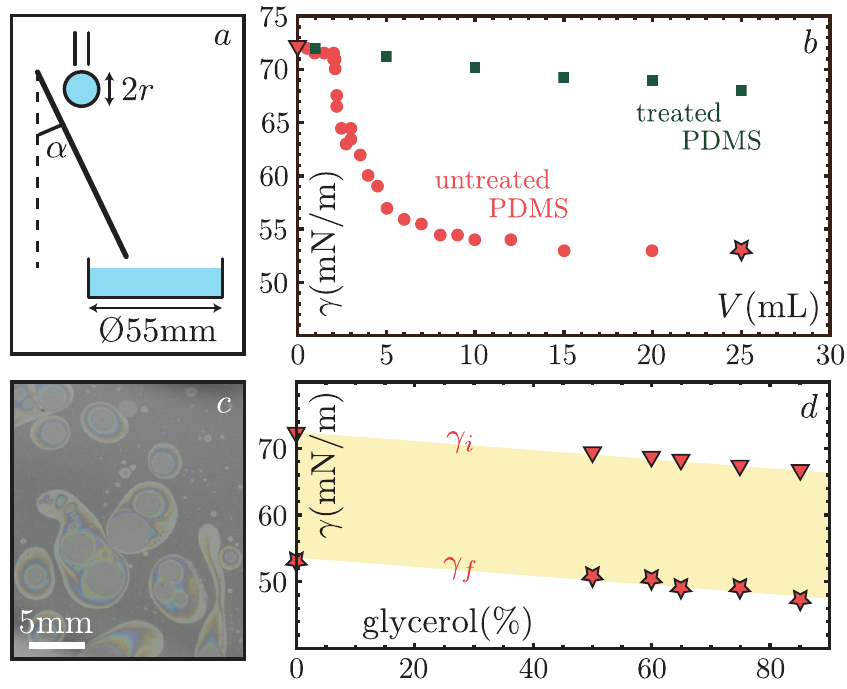}
\caption{(a)~Experimental setup, with $\alpha=29.2^{\circ}$ and $r=2$~mm. (b)~Surface tension of given volumes of water droplets collected just after reaching the second speed regime during their descent on untreated PDMS (red circles), compared to the same experiment performed on toluene-treated PDMS, with droplets collected after their descent on the 10~cm sample (dark green squares). (c)~Surface of a beaker in which 1500 water droplets (of 33~$\mu$L each) were collected after reaching the second speed regime. The colored zones correspond to thicknesses of the order of light wavelengths, and transparent lenses correspond to thicker oil zones. (d)~Initial and final values of the surface tension for the different liquids described in Table \ref{tbl:liquids}. Error bars are of the order of magnitude of the markers size.} 
\label{fig:surfacetension}
\end{figure}

Observing a sudden decrease of surface tension in our experiment would thus bring additional information about the presence of a thin oil film at the droplet-air interface: we focus on the surface tension of the droplets before being deposited on the substrate and after reaching the second regime. An in-situ measurement of the droplet-air surface tension proves to be difficult, and hanging droplet measurements have the drawback of requiring droplet collection inside a capillary tube, possibly breaking a thin oil film present on the droplets.
Thus, we use the following setup: water-glycerol droplets of radius $r=2.0$ mm are deposited with a syringe pump onto an inclined PDMS plane (Fig.~\ref{fig:surfacetension}a). 
%
Droplets are rolling down the plane and collected in a 55mm diameter Petri dish after reaching the second regime.
%
Surface tension of the liquid in the Petri dish is then measured as function of the collected volume for both droplets rolling down untreated Sylgard 184 PDMS and toluene-treated PDMS.
For the untreated sample, Fig.~\ref{fig:surfacetension}b shows a dramatic decrease of surface tension around a collected volume of 3 mL, which corresponds to about 100 droplets. 
If we assume that each droplet is covered by a homogeneous thin film of oil, we calculate that the surface covered by this liquid oil in the Petri dish would be about 100 times larger than on one single droplet. A collected volume of 3 mL thus corresponds to the situation for which the thickness of the oil film on the Petri dish is of the same order as the thickness of the oil film on a droplet in the second regime. The same experiment performed with various Petri dish sizes and materials (polystyrene and glass) results in a collapse of all the data when rescaled with the Petri dish surface, as shown in Fig. S3, Supplementary Information.
We therefore conjecture that as the droplet is rolling down on PDMS, it is gradually covered with oil and eventually submitted to a change in surface tension. Building on results from Lee \cite{Lee1991}, we argue that the surface tension change occurs when a critical surface concentration on the droplet is reached, corresponding to the presence of a homogeneous oil film. We infer that this sharp transition in terms of surface tension is linked to a sudden change in the droplet state: before the transition, the droplet is only partially covered with some patches of uncrosslinked oligomers, while after the transition, the droplet is completely covered by an oil layer.
Hence, we explain the sudden change in speed between the two regimes by the sudden transition between these two states for the droplet. 

We also generalize the results obtained for water to various water-glycerol mixing ratios. The initial surface tension $\gamma_i$ in the Petri dish before droplet collection, and the final surface tension $\gamma_f$ measured when a 25~mL volume of droplets has been collected are both shown in Fig.~\ref{fig:surfacetension}d as a function of water-glycerol mixing ratio.
We see that the difference between the initial and final surface tensions is constant, $\gamma_i-\gamma_f\simeq 19$~mN/m, showing that the surface pressure of the oil film formed at the surface of the liquid bath when collecting 25~mL of droplets remains constant as a function of the water-glycerol mixing ratio.

%
%
%
%
%
%
%
%
\section{Quantitative speed measurements and discussion}
%
%
%
%
%
%
%
%
%
%
\subsection{Experimental observations}
\begin{figure}[bht]
\noindent\includegraphics[width=8.7cm]{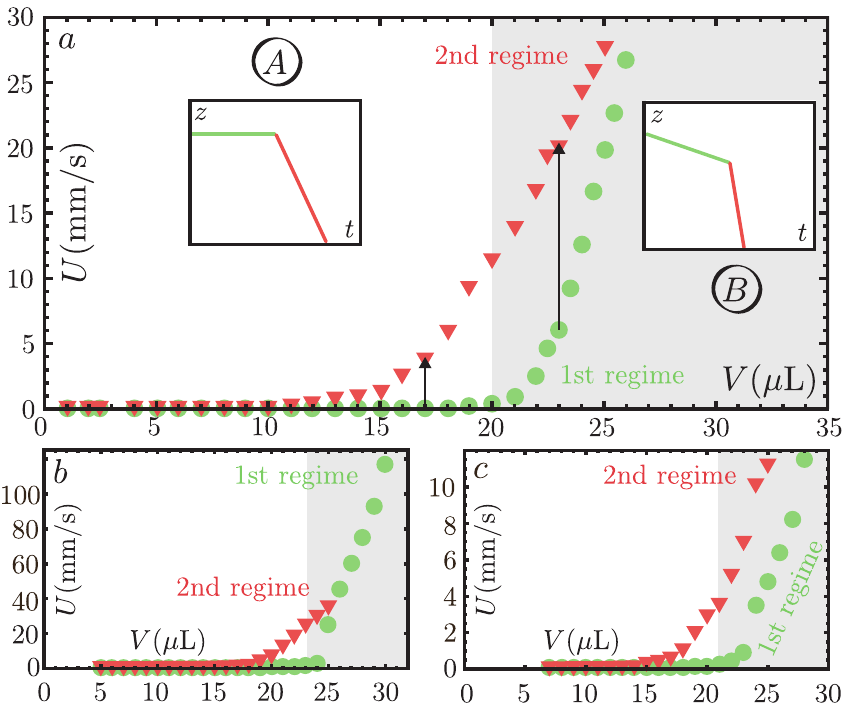}
\caption{(a)~65\% glycerol - 35\% water droplet speeds on untreated PDMS as a function of its volume, in first (light green circles) and second regime (red triangles). Insets: schematic phase diagrams in zone A (the speed in the first regime is negligible compared to the speed in the second regime) and zone B (the speed in the first regime is not negligible compared to the speed in the second regime). (b)~Same experiment for a 50\% glycerol - 50\% water mixture. (c)~Same experiment for a 75\% glycerol - 25\% water mixture. Error bars are of the order of magnitude of the markers size.}
\label{fig:speeds1}
\end{figure}

%
A quantitative measurement of the velocities in the first and second regimes on untreated PDMS is shown in Fig.~\ref{fig:speeds1}a as a function of droplet volume, for a 35\% water and 65\% glycerol mixture. The difference between the first and second regime speed is measured by looking at a vertical line drawn on this graph (black arrows).
The graph can be divided into two zones A and B. In zone A, the speed in the first regime is negligible compared to the speed in the second regime. As shown on the height versus time diagrams in inset of Fig.~\ref{fig:speeds1}a, in zone A, a droplet deposited on a vertical PDMS substrate seems to be unmoving until reaching the second regime once the droplet is covered by oil. In zone B, the speed in the first regime is not negligible compared to the speed in the second regime.
%

%

Examples for different water-glycerol mixing ratios are given in Fig.~\ref{fig:speeds1}b and c. In Fig.~\ref{fig:speeds1}b, we show results obtained for a less viscous fluid, with a 50\% glycerol - 50\% water mixture \footnote{For this mixture, only one regime is observed for volumes larger than 25 $\mu$L. Corresponding points are classified as first regime measurements. The same behavior is also observed (for higher volumes) for higher glycerol/water mixing ratios. Although our understanding of this phenomenon is not complete, an hypothesis could be that a droplet moving too fast does not have time to get covered by uncrosslinked chains.}. In Fig.~\ref{fig:speeds1}c, we show results obtained for a more viscous fluid, with a 75\% glycerol - 25\% water mixture.
By comparing Fig.~\ref{fig:speeds1}a, b and c, we observe that the difference in behavior between the first and second regime (that can be quantified by looking at the shape of the experimental curves at large volumes) depends on the fluid viscosity. This phenomenon will be further discussed in the following sections.

Another experimental observation is that in the first regime, there is a volume under which the speed of the droplet is zero \footnote{\textit{Zero} being defined as moving by less than a few pixels on the recorded video during the whole duration of the first regime.} ({\em e.g.} 11~$\mu$L in the experiments presented in Fig.~\ref{fig:speeds1}a, although first regime speeds values remain negligible compared to second regime speeds until 20~$\mu$L) whereas in the second regime, the speed of the smallest deposited droplets is not exactly zero: there seems to be no threshold volume below which a droplet does not move in the second regime.

\subsection{Droplet speed in the first regime}

In this discussion paragraph, we compare the dynamics in the first regime on untreated PDMS to the single regime dynamics on toluene-treated PDMS.

\begin{figure}[bht]
\noindent\includegraphics[width=8.7cm]{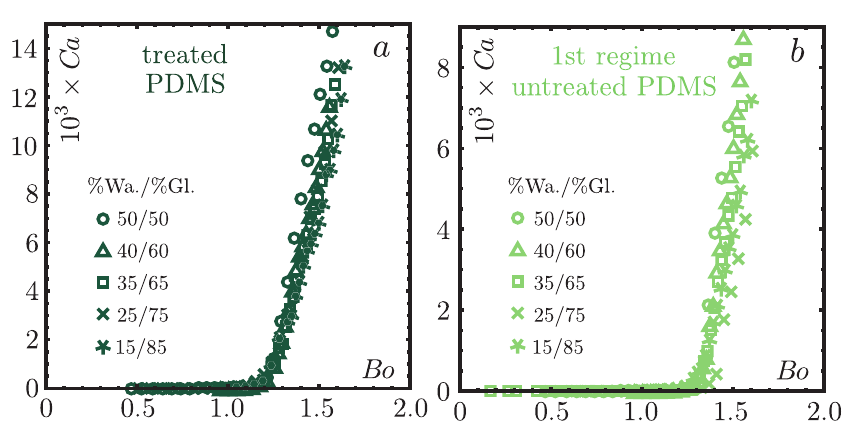}
\caption{Capillary number $\Capp$ as a function of Bond number $\Bond$, for all liquids described in Table \ref{tbl:liquids}. (a) corresponds to results on treated PDMS and (b) to results for the speed in the first regime on untreated PDMS. Error bars are of the order of magnitude of the markers size.}
\label{fig:CavsBo}
\end{figure}

The speeds measured in the single regime observed on a toluene-treated PDMS sample are shown in Fig.~\ref{fig:CavsBo}a. All results of the experiments conducted with the different liquids mentioned in Table \ref{tbl:liquids} collapse onto a single curve when using the dimensionless numbers $\Capp$ and $\Bond$: below a critical $\Bond$ number, the droplet  is virtually immobile as pinning forces overcome the droplet weight. Above this critical $\Bond$ number, the relationship between $\Capp$ and $\Bond$ is linear. Results obtained in the first regime on an untreated PDMS sample are shown with the same data processing, and a collapse onto a single curve is also obtained when plotting $\Capp$ as a function of $\Bond$. This result means that in the first regime, the dynamics is set by the usual balance between gravity, capillary pinning forces, and viscous dissipation inside the droplet. Our results in the first speed regime are consistent with the first order theory derived at the beginning of section 3.

\begin{figure}[bht]
\noindent\includegraphics[width=8.7cm]{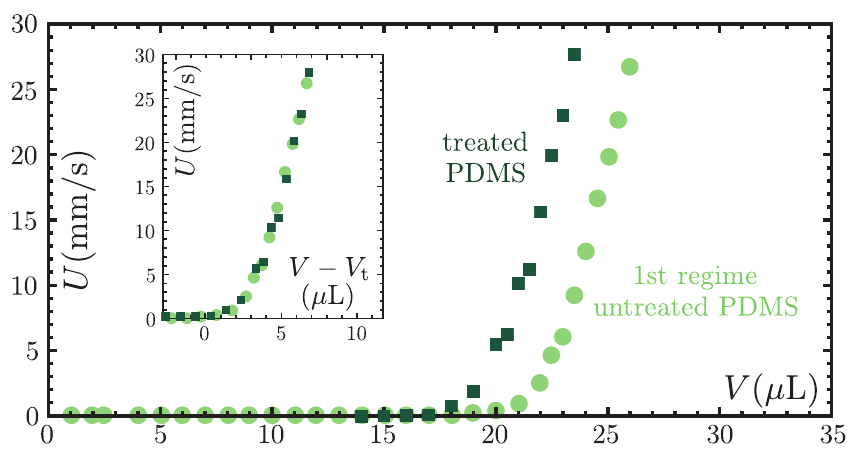}
\caption{65\% glycerol - 35\% water droplet speed on treated PDMS (dark green squares) as a function of its volume, compared to the results for the first speed on untreated PDMS (light green circles). {\em Inset}: speed as a function of the difference between the droplet volume $V$ and the threshold volume $V_t$, on treated PDMS (dark green squares) and on untreated PDMS (light green circles). Error bars are of the order of magnitude of the markers size.}
\label{fig:PDMSvstoluene}
\end{figure}

A comparison between measurements on treated samples in the first regime and on untreated samples is performed in Fig.~\ref{fig:PDMSvstoluene}, for a 35\% water and 65\% glycerol mixture. For clarity sake, we choose to do the comparison on only one water glycerol mixing ratio without loss of generality. The shape of the two curves obtained on treated PDMS and on untreated PDMS is the same, except for the threshold volume value $V_t$ below which the droplet speed is negligible (Inset of Fig.~\ref{fig:PDMSvstoluene}). 
The small difference in the threshold volumes $V_t$ between the two curves can be explained by differences in advancing and receding angles on the two different samples \footnote{The values measured for advancing and receding contact angles are compatible with this hypothesis, but the uncertainties on our measurements of the contact angle hysteresis are too large to use these values to rescale our data.}. Hence we conclude that a droplet in the first regime on untreated PDMS behaves exactly as a droplet on a treated sample, i.e. as if it was not polluted by uncrosslinked oligomers. Some uncrosslinked oligomers are certainly present on the droplet but do not affect its dynamics. Especially, uncrosslinked oligomers might be present preferentially at the surface of water, forming no significant oil wetting ridges during the first regime. Oil wetting ridges are likely to both lubricate the droplet and dissipate energy, causing a different dynamics.

\subsection{Droplet speed in the second regime}

The droplet dynamics on untreated PDMS in the second regime involves two main differences with the dynamics in the first regime: there seems to be no threshold volume below which a droplet does not move, and the dissipation processes appear to occur both in the droplet and in the oil layer. 

\begin{figure}[bht]
\noindent\includegraphics[width=8.7cm]{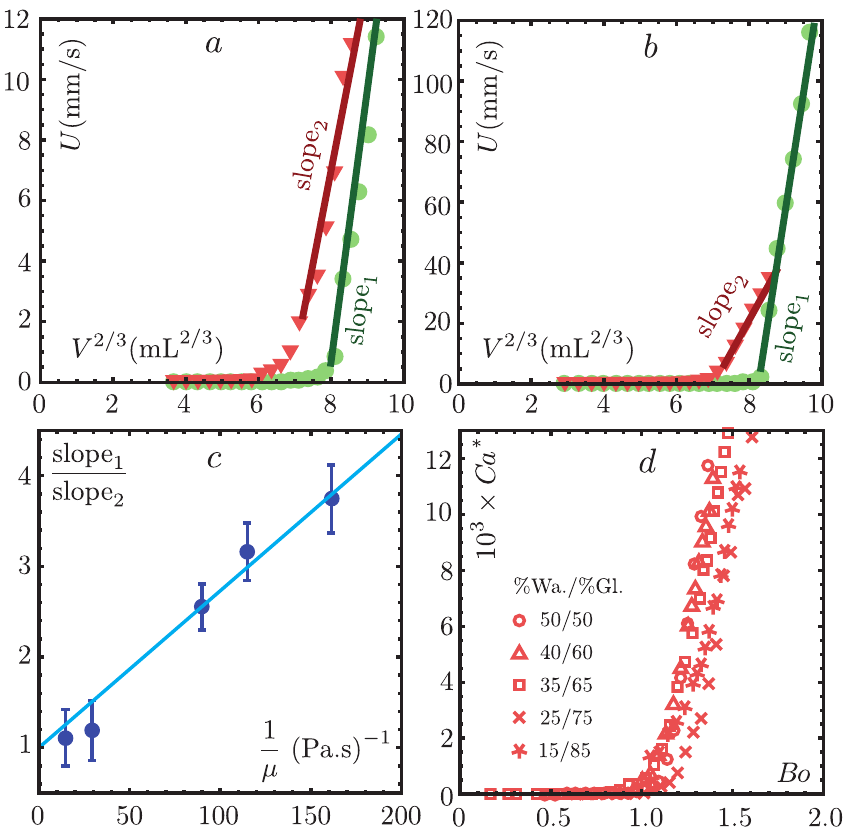}
\caption{(a)~Droplet speed in the first (green circles) and second (red triangles) regimes in the case of a 75\% glycerol - 25\% water mixture, as a function of $V^{2/3}$, where $V$ is the droplet volume. The linear part of the curve is fitted and the slope of the fit is denoted slope$_1$ for the first regime and slope$_2$ for the second regime. (b)~Same graphic for a 50\% glycerol - 50\% water mixture. (c)~Ratio between the slopes measured for the linear part in the first and second regimes, as a function of the inverse of the water-glycerol mixture viscosity. An affine relationship with intercept equal to unity is found, in agreement with our simple model. (d)~A modified Capillary number $\Capp^*$ is used to plot the speed of the droplets in the second regime as a function of their weight, through the $\Bond$ number.}
\label{fig:slopes}
\end{figure}

Indeed, the asymptotes of the curves in the first and second regime in Fig.~\ref{fig:slopes}a and b are almost parallel for a viscous fluid (Fig.~\ref{fig:slopes}a) while for a less viscous liquid, the slope in the second regime  is smaller than the slope in the first regime (Fig.~\ref{fig:slopes}b). This is an indication for some dissipation occurring inside the oil cover, visible in the case of a less viscous droplet, but hidden by the huge dissipation rate in the droplet in the case of a viscous droplet. The contribution of the oil cover to the dissipation could come both from the oil thin layer wrapping the droplet and from oil wetting ridges. Given the estimated thickness for the oil layer, we assume that the dissipation in the oil occurs mainly in wetting ridges.

In the literature, the dynamics of a water droplet on a Slippery Lubricant Infused Surface is shown to depend on the oil viscosity. Smith et al \cite{Smith2013} evaluate the contribution of each viscous dissipation term (viscous dissipation in the droplet, viscous dissipation in the thin oil film below the droplet, viscous dissipation in the oil ridges). They conclude that taking into account only the dissipation rate inside the wetting ridge, which dominates over the two other terms, is sufficient to explain their experimental data. However, a rescaling using the model proposed by Smith et al does not lead to a collapse of all the data in our experiments: in the case of a viscous droplet, the dissipation inside the droplet is not smaller than the dissipation in the oil wetting ridge.

Here we propose a simple model involving two dissipation terms: viscous dissipation in the droplet and viscous dissipation in the oil wetting ridge. The viscous dissipation in the oil ridge is assumed to be equal for droplets of the same size but of various water-glycerol mixing ratios. The viscous force that brakes the droplet is then not only
\begin{equation}
F_{\mu}= A \frac{\mu U \mathcal{S}}{h} \simeq A \mu U V^{1/3}
\end{equation}
but 
\begin{equation}
F_{\mu\textrm{tot}}= A \mu U V^{1/3}+ B \mu_o U R_b
\end{equation}
where $\mu_o$ is the PDMS oil viscosity, $R_b\simeq V^{1/3}$ is the radius of the contact zone of the droplet on the substrate\cite{Smith2013}, and A and B are two constants linked to the geometry of the droplet and of the wetting ridge, assumed to be independent to the droplet size and water-glycerol mixing ratio to keep the model as simple as possible.

If using the theoretical framework described at the beginning of section 3, the slope of the asymptotes at large volume in Fig.~\ref{fig:slopes}a and b should be proportional to $1/(A\mu)$ in the first regime (slope$_1$), and to $1/(A\mu+B\mu_o)$ in the second regime (slope$_2$). 
In this case, the relationship between slope$_1$, slope$_2$, $A$, $B$, $\mu$, and $\mu_o$ yields:

\begin{equation}
\frac{\textrm{slope}_1}{\textrm{slope}_2}=1+\frac{B\mu_o}{A\mu}
\end{equation}

We plot in Fig. \ref{fig:slopes}c the ratio between the two slopes as a function of 1/$\mu$. The results are consistent with an affine relationship, with an intercept equal to unity. This curves gives us information about the ratio of energy dissipated in the droplet and in the wetting ridge: the slope of the affine fit is equal to $C=B\mu_o/A=0.0173$.
To plot all our data on a same graph, we then choose a new version of the $\Capp$ number, defined as:
\begin{equation}
\Capp^*=(\mu+C)U/\gamma
\end{equation}

\noindent This allows all the curves to have the same slope for the linear part at high droplet volumes (Fig. \ref{fig:slopes}d).

However, an experimental observation is not taken into account in this model: we do not observe a threshold volume below which a droplet does not move in the second regime, but there seems nevertheless to be a threshold volume below which the droplet speed is negligible.
This phenomenon might be linked to a dependence of the contact angle hysteresis on the geometrical properties of both the droplet and the wetting ridge. A newly published study \cite{Semprebon2017} performed in the case of liquid infused surfaces highlights the dependence of contact angle hysteresis of a droplet surrounded by an oil wetting ridge on the ratio between the Laplace pressures inside the droplet and inside the wetting ridge. Although a priori valid only in the case where the droplet is not cloaked by oil, these results interestingly suggest a candidate phenomenon to explain an apparent variation of the pinning force as a function of the droplet volume in our experiments.


\section{Conclusions}
We have shown that uncrosslinked chains found in most commercial elastomers are responsible for an unexpected droplet dynamics: an aqueous droplet deposited on a vertical silicone elastomer plate exhibits successively two  different regimes with two different constant speeds. 
This phenomenon disappears if the elastomer is treated in order to extract  uncrosslinked chains from the silicone network, and reappears when re-swelling such an elastomer with a silicone oil, thus demonstrating the crucial role played by uncrosslinked chains in the droplet dynamics. 
Our study reveals how minute amounts of contaminates have dramatic effects on the wetting dynamics.
A direct vizualisation of the oligomers has been performed at the surface of a liquid bath composed of thousands of droplets collected after their descent on untreated PDMS samples. 
We have also shown that the sudden change observed in droplets speeds coincides with a sudden change in surface tension due to the surface contamination by silicone oligomers: we conjecture that before this surface tension change, the droplet is only partially covered with some patches of uncrosslinked chains, while after the surface tension decrease, a thin oil film completely covers the droplet. The dynamics in the first regime is different from the dynamics in the second regime because the droplet exhibits two different states: in the first regime the droplet can be modeled by a pure liquid droplet, with no impact of the uncrosslinked oil chains on its dynamics, while in the second regime, the oil cover (including possible wetting ridges) has to be taken into account to explain the droplet dynamics.
Our findings could impact various research domains such as microfluidics or elastocapillarity as it contributes to a better knowledge of the interaction between water and silicone elastomers, and provides a simple test to evaluate the presence of unintended free oligomer chains by looking at the dynamics of water droplets on a test surface.

\section*{Acknowledgements}

We thank Laurent Limat, C{\'e}dric Boissi\`ere, Paul Grandgeorge, and Alexis Prevost for discussions.
The present work was supported by ANR grant ANR-14-CE07-0023-01.

%

%


\bibliographystyle{apsrev4-1}
\bibliography{DropletsDynamicsOnElastomers}

\end{document}